\documentclass{article}
\usepackage{spconf,amsmath,graphicx}

\usepackage[T1]{fontenc}
\usepackage[latin9]{inputenc}
\usepackage{float}
\usepackage{calc}
\usepackage{amsmath}
\usepackage{graphicx}
\usepackage{subcaption}
\usepackage{mathtools}
\usepackage[export]{adjustbox}
\usepackage{listings}
\usepackage{amssymb}
\usepackage{flexisym}
\usepackage{xcolor}
\usepackage{wrapfig}
\usepackage{textcase}
\usepackage{soul}
\usepackage{comment}
\usepackage{multirow}

 \graphicspath{{./imgs/}}
 \usepackage[ruled,vlined]{algorithm2e}
\usepackage{tabularx,booktabs}
\usepackage[numbers,sort&compress]{natbib}
\usepackage{tikz}
\usetikzlibrary{positioning}

\title{Ringing Artifact Reduction Method for Ultrasound Reconstruction using Multi-Agent Consensus Equilibrium}

\name{%
\begin{tabular}{@{}c@{}}
Abdulrahman M. Alanazi$^{1,3}$ \qquad 
Singanallur Venkatakrishnan$^{2}$ \qquad 
Gregery T. Buzzard$^{1}$ \\
Charles A. Bouman$^{1}$ 
\end{tabular}}
 \address{$^{1}$ Purdue University-Main Campus, West Lafayette, IN 47907.  \\
     $^{2}$ Oak Ridge National Laboratory, One Bethel Valley Road, Oak Ridge, TN 37831.\thanks{This manuscript has been supported by UT-Battelle, LLC under Contract No. DE-AC05-00OR22725 with the U.S. Department of Energy. G. Buzzard was partially supported by NSF CCF-1763896, and C. Bouman was partially supported by the Showalter Trust. The United States Government retains and the publisher, by accepting the article for publication, acknowledges that the United States Government retains a non- exclusive, paid-up, irrevocable, world-wide license to publish or reproduce the published form of this manuscript, or allow others to do so, for United States Government purposes. The Department of Energy will provide public access to these results of federally sponsored research in accordance with the DOE Public Access Plan (http://energy.gov/downloads/doe-public-access-plan).} \\
     $^{3}$King Saud University (KSU), Riyadh, Saudi Arabia.}
\begin{document}
\ninept
\maketitle
\begin{abstract}
Non-destructive characterization of multi-layered structures that can be accessed from only a single side is important for applications such as well-bore integrity inspection. 
Existing methods related to Synthetic Aperture Focusing Technique (SAFT) rapidly produce acceptable results but with significant artifacts.  
Recently, ultrasound model-based iterative reconstruction (UMBIR) approaches have shown significant improvements over SAFT. 
However, even these methods produce ringing artifacts due to the high fractional-bandwidth of the excitation signal. 

In this paper, we propose a ringing artifact reduction method for ultrasound image reconstruction that uses a multi-agent consensus equilibrium (RARE-MACE) framework. 
Our approach integrates a physics-based forward model that accounts for the propagation of a collimated ultrasonic beam in multi-layered media, a spatially varying image prior, and a denoiser designed to suppress the ringing artifacts that are characteristic of reconstructions from high-fractional bandwidth ultrasound sensor data.
We test our method on simulated and experimental measurements and show substantial improvements in image quality compared to SAFT and UMBIR.
\end{abstract}
\section{Introduction}
\label{sec:intro}

Non-destructive evaluation (NDE) of multi-layered structures that can be accessed from only a single side is important in many applications.
For example, this imaging scenario occurs when monitoring the structural integrity of oil and geothermal wells that lie behind layers of fluid and steel casing.
While ultrasound imaging is widely used in NDE applications, multi-layered structures present a challenge for ultrasounds systems because of the complex propagation and reverberation of the signal through the material.

The most popular methods to reconstruct data from ultrasound systems use a  delay-and-sum (DAS) approach because of their low computational complexity. 
One such approach is the synthetic aperture focusing technique (SAFT), which produces acceptable ultrasound images for simple objects \cite{prine1972synthetic}. 
SAFT has been applied to single-layer \cite{stepinski2007implementation,hoegh2015extended} and multi-layered structures \cite{skjelvareid2011synthetic,lin2018ultrasonic} but not to collimated beam systems. 
Nevertheless, SAFT and its variations rely on a simple model that can lead to ringing artifacts and blur.

In order to reduce the reconstruction artifacts of SAFT while maintaining computational efficiency,  regularized inversion can be used with a linear propagation model.
In \cite{ozkan2017inverse}, the forward model is extended to handle plane-wave imaging.  
Recently, the ultrasound model-based iterative reconstruction (UMBIR) approach of \cite{almansouri2018model} used a propagation model of the ultrasound through the medium and combined all the data from the source-detector pairs to jointly reconstruct a fully 3D image.
Most recently, UMBIR was upgraded to account for collimated beam systems and multi-layer structures in \cite{alanazi2022model}. 
However, even these methods exhibit ringing artifacts due to the high fractional-bandwidth excitation signal. 
One way to overcome this issue is to jointly process multi-frequency data sets as presented in \cite{alanazi2022modelX}. This method can significantly reduce ringing artifacts but only when multi-frequency data sets are available.

In this paper, we propose a framework that we refer to as ringing artifact reduction using multi-agent consensus equilibrium (RARE-MACE) designed to reduce ringing artifacts in ultrasound images. 
Our algorithm integrates three distinct agents using the MACE framework \cite{buzzard2018plug,sreehari2017multi,majee20194d,9143147,bouman2022foundations}. 
The core contribution of this work is the introduction of a ringing artifact reduction agent based on the recently proposed BM3D extension that incorporates a spatially correlated noise kernel~\cite{makinen2020collaborative}.
This new agent is combined using the MACE framework with a physics-based forward model and a spatially varying  q-generalized Gaussian Markov random field (QGGMR) model~\cite{alanazi2022model,alanazi2022modelX} to compute the final reconstruction.
Our experiments using both simulated and experimental measurements demonstrate that our proposed RARE-MACE framework substantially suppresses ringing artifacts compared to SAFT and UMBIR. 

In Section~\ref{sec:RingingArtifactsAnalysis}, we briefly analyze ringing artifacts in ultrasound image reconstructions.
In Section~\ref{sec:ProblemFormulation}, we formulate the problem, describe the MACE framework, and introduce our ringing artifacts agent. 
Finally, we present our experimental results in Section~\ref{sec:results} and draw conclusions in Section~\ref{sec:conc}.

\vspace*{-\baselineskip}
\section{Ringing Artifacts Analysis}
\vspace*{-\baselineskip}
\label{sec:RingingArtifactsAnalysis}

\begin{figure}[htb]
\centering
\includegraphics[width=0.4\textwidth]{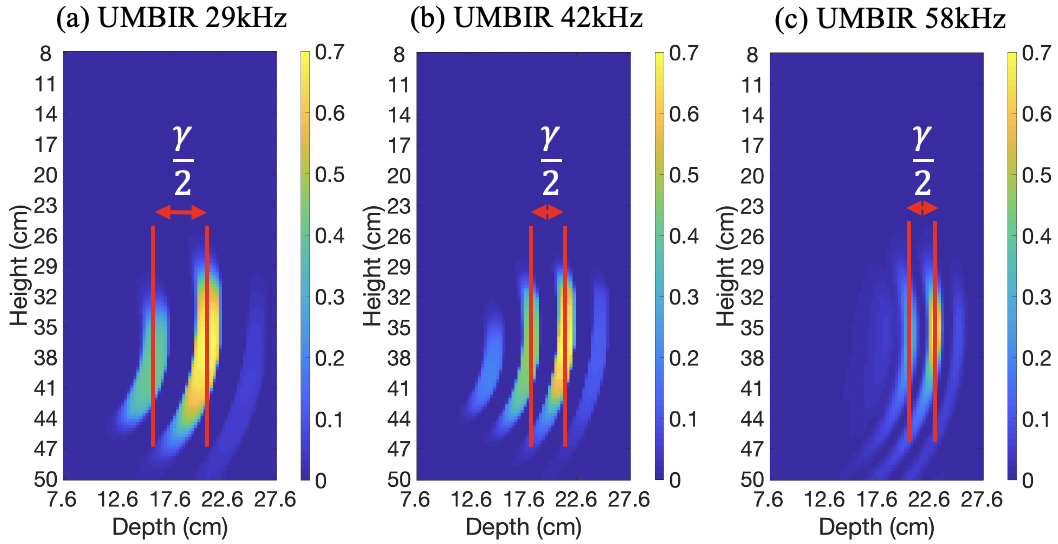}
\caption{UMBIR reconstructions from our previous work in \cite{alanazi2022modelX} associated with a center frequency, $f_c$, of (a) $29 kHz$, (b) $42 kHz$, and (c) $58 kHz$. Note that the ringing artifacts are separated by $\frac{\gamma}{2}$. }
\label{fig:reverb_analysis}
\end{figure}

When an acoustic excitation signal which has a high fractional bandwidth is used in ultrasound systems, as is the case in~\cite{alanazi2022model,alanazi2022modelX}, then the associated reconstructions tend to have characteristic ringing artifacts associated with the center frequency of the signal.
In particular, we define the following characteristic wavelength given by
\begin{eqnarray}
\label{eq:spatial-frequency-reverb}
\gamma = \frac{ c_m }{ f_c \Delta_p },
\end{eqnarray}
where $c_m$ is the velocity of sound in the medium in $\frac{m}{s}$, $f_c$ is the center frequency of the excitation signal in $Hz$, and $\Delta_p$ is the pixel pitch (i.e., spacing) in $m$. 
Then we have observed empirically that ringing artifacts in reconstructions will have a characteristic period of approximately $\gamma/2$.
Figure~\ref{fig:reverb_analysis} shows some typical examples of UMBIR reconstructions using the data reported in~\cite{alanazi2022modelX} for three different center frequencies of $29 kHz$, $42 kHz$, and $58 kHz$ with acoustic velocity of $c_m = 2620 \frac{m}{s}$, and voxel sample spacing of~$\Delta_p = 3 cm$.
Note that the ringing artifacts from the reflection have a fixed pattern with peaks separated by $\frac{\gamma}{2}$.
In Section~\ref{sec:ReverberationReductionAgents}, we will use the value of $\gamma$ from~\eqref{eq:spatial-frequency-reverb} in order to design an algorithm  
that effectively rejects these artifacts from the space of possible reconstructions.

\vspace*{-\baselineskip}
\section{Problem Formulation}
\vspace*{-\baselineskip}
\label{sec:ProblemFormulation}
Assuming a linear system for simplicity, we seek to reconstruct an image $x$ using a measurement model of the form 
\begin{equation}
    \label{eq:sysmodel}
    y = Ax + Dg + e, 
\end{equation}
where $y\in \mathbb{R}^{MK \times 1}$ is a vector of measurements from $K$ receivers at $M$ timepoints, $A \in \mathbb{R}^{MK\times N}$ is the system matrix, $x\in \mathbb{R}^{N \times 1}$ is the vectorized version of the desired image with $N$ voxels, $D\in \mathbb{R}^{MK \times K}$ is a matrix whose columns form a basis for the possible direct arrival signals, $g \in \mathbb{R}^{K \times 1}$ is a scaling coefficient vector for $D$, and $e$ is a Gaussian random vector with distribution $N(0, \sigma^2 I)$. 
Details for constructing the system matrices, $A$ and $D$, can be found in \cite{alanazi2022model,alanazi2022modelX}.

\vspace*{-\baselineskip}
\subsection{MACE Framework}
\label{sec:mace_framework}
In order to find find a solution to the inverse problem corresponding to \eqref{eq:sysmodel}, we use the multi-agent consensus equilibrium (MACE) framework \cite{buzzard2018plug}. 
MACE finds a solution that achieves an equilibrium between multiple agents, each enforcing certain desirable characteristics in the reconstruction.  
We use three agents in our MACE formulation. 
The first agent is a proximal map forward model agent that promotes data fidelity. 
The other two agents are denoisers designed to suppress ringing artifacts and promote spatial regularity. 


In order to formulate the MACE equations, we define $W \in \mathbb{R}^{3\times N}$ to be a stack of $3$ independent states, where each state is the input to its corresponding agent.  Hence $W$ has the form
\begin{align}
\label{eq:stacked_states}
W = 
\left[ 
\begin{array}{c}
w_0  \\
w_1  \\
w_2 
\end{array}
\right] \ . 
\end{align}
Next, we define the agent operator as 
\begin{align}
\label{eq:agent_operator}
L(W) = 
\left[ 
\begin{array}{c}
F(w_0)  \\
H_1(w_1)  \\
H_2(w_2) 
\end{array}
\right] ,
\end{align}
where $F$ is the forward model agent, and $H_1$ and $H_2$ are the denoising agents used to suppress artifacts. 
Note that \eqref{eq:agent_operator} simply denotes the simultaneous application of each agent on its respective state vector. 
We then define the averaging operator 
\begin{align}
\label{eq:average_operator}
G(W) = 
\left[ 
\begin{array}{c}
\Bar{w}  \\
\Bar{w}  \\
\Bar{w} 
\end{array}
\right]
\end{align}
using a weighted average defined as 
\begin{eqnarray}
\label{eq:weighted_average}
\Bar{w} = \frac{1}{1+\mu} w_0 + \frac{\mu}{2(1+\mu)} \left( w_1 + w_2 \right) ,
\end{eqnarray} 
where the unitless parameter $\mu$ controls the strength of regularization versus data fitting.
Note equal weight between the forward agent and two prior agents can be achieved by setting $\mu=1$.
With this framework, the consensus equilibrium solution as in  \cite{buzzard2018plug,9143147} is $W^*$ that satisfies 
\begin{eqnarray}
\label{eq:CE_equation}
L(W^*) = G(W^*)  \ .
\end{eqnarray}
In this case, the MACE solution is given by $\hat{x} = \Bar{w}^*$, where $\Bar{w}^*$ is the average of the stacked components in $W^*$. 
It is shown in \cite{bouman2022foundations,buzzard2018plug} that the consensus equilibrium equations can be solved using the Douglas-Rachford (DR) algorithm with the update 
\begin{eqnarray}
\label{eq:DR_update}
W^{i+1} = W^i + \rho (T W^i - W^i) , 
\end{eqnarray}
where $\rho \in (0,1)$ controls the speed of convergence, and the solution is exactly the fixed point of the operator $T = (2G-I)(2L-I)$.
The pseudocode of MACE for practical implementation is available in \cite{bouman2022foundations}. 

The data fitting agent is defined as 
\begin{eqnarray}
\begin{aligned}
F(x) = {\text{arg}} \min_z \min_g \biggl\lbrace \frac{1}{2\sigma^2} \lVert  y-Az-Dg\rVert^{2}  + \frac{1}{2\beta} \lVert z - x \rVert^2  \biggl\rbrace ,
\end{aligned}
\label{eq:first_agent}
\end{eqnarray}
where $\beta >0$ controls the convergence speed.
As in \cite{alanazi2022model,bouman2022foundations}, we solve \eqref{eq:first_agent} using Iterative Coordinate Descent.
For the second agent, $H_1$, we adopt the spatially varying q-generalized Gaussian Markov random field (QGGMRF) based denoiser that preserves both low-contrast characteristics and edges from \cite{alanazi2022model,thibault2007three}. 
In the next subsection we describe our core contribution, which is the use of a specially designed BM3D agent for $H_{2}$ in order to suppress ringing artifacts. 

\vspace*{-\baselineskip}
\subsection{The Ringing Artifacts Reduction Agent}
\label{sec:ReverberationReductionAgents}

In this section, we describe the agent, $H_2$, that we use to suppress the ringing artifacts in the reconstructed image caused by the high fractional-bandwidth excitation signal.
To do this, we use the recently published version of BM3D in \cite{makinen2020collaborative}.
Unlike the standard BM3D, this new version allows the use of a spatially correlated noise kernel to model fixed-pattern noise such as the ringing artifacts that can occur in ultrasound reconstructions. 

For our application, we modify the correlated noise kernel with a circular repeating pattern \cite{makinen2020collaborative} to be
\begin{eqnarray}
\label{eq:NoiseKernel}
P = \text{cos} \left( \frac{\sqrt{\left( x^{(1)}\right)^2 + \left( x^{(2)}\right)^2}}{ \gamma} \right) G_{\eta} \left( x^{(1)}, x^{(2)} \right) ,
\end{eqnarray}
where $x^{(1)}$ and $x^{(2)}$ denote the horizontal and vertical coordinates in units of samples, respectively, and $G_{\eta}$ is a 2-D Gaussian function centered at the origin, with a standard deviation of $\eta$.
The additional parameter $\gamma>0$ we added is a scaling parameter that describes the spatial frequency of the ringing artifacts as defined in~\eqref{eq:spatial-frequency-reverb}.  

\begin{figure}[htb]
\centering
\includegraphics[width=0.4\textwidth]{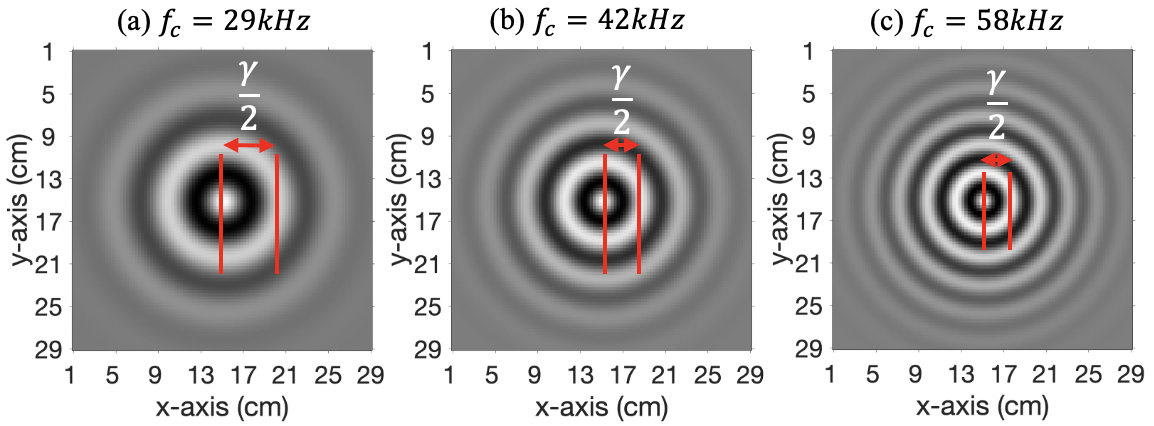}
\caption{A visualization for the spatially correlated noise kernel in \eqref{eq:NoiseKernel} with $\eta = 20$, $c_m = 2620 \frac{m}{s}, \Delta_p = 3 cm$, center frequency, $f_c$, of (a) $29kHz$, (b) $42kHz$, and (c) $58kHz$. The rings are separated by $\frac{\gamma}{2}$.
}
\label{fig:kernel}
\end{figure}
Figure~\ref{fig:kernel} illustrates the noise kernel in \eqref{eq:NoiseKernel} using different center frequencies. 
The parameters used are $\eta = 20$, $c_m = 2620 \frac{m}{s}$, $\Delta_p = 3 cm$, $f_c = $ $29kHz$, $42kHz$, and $58kHz$. 
Note that the spacing between the rings is fixed at $\frac{\gamma}{2}$, which matches the spacing between the ringing artifacts in Figure~\ref{fig:reverb_analysis}. 
In the next section, we will show how incorporating BM3D with this designed spatially correlated noise kernel into MACE reconstruction is effective for reducing ringing artifacts.

\vspace*{-\baselineskip}
\section{Experimental Results}
\label{sec:results}

\begin{figure}[htb]
\centering
\includegraphics[width=0.4\textwidth]{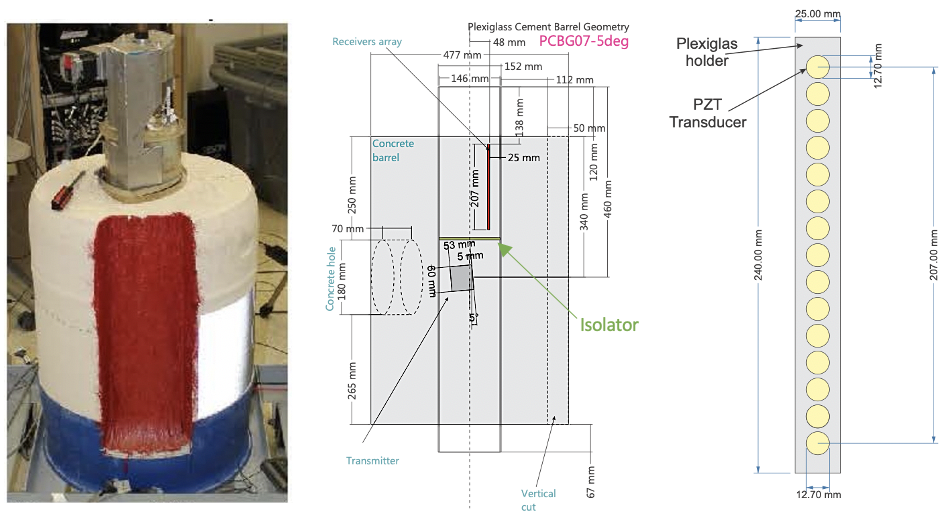}
\caption{Left: A picture of the specimen to be imaged using the collimated beam system.  The red region is a notch that is obtained by cutting a section of the cylinder. Middle: The system geometry. Right: the receiver geometry.}
\label{fig:LANL_I_sys}
\end{figure}

In this section, we compare the proposed RARE-MACE method against UMBIR and SAFT using synthetic and real data sets and also demonstrate the effect of the BM3D agent on reconstructions. 
The synthetic data was generated using the K-Wave simulation package \cite{treeby2010k}. 
Both our synthetic and measured data experiments are designed to evaluate the performance of the well-bore integrity inspection system designed at the Los Alamos National Laboratory (LANL) and shown in Figure~\ref{fig:LANL_I_sys}.
This system uses a collimated acoustic transmitter along with an array of 15 receiving transducers.
The sensor assembly is embedded in a water-filled borehole in the center of a concrete cylinder. 
The borehole is lined with a thin layer of Plexiglas that holds the entire system. 
The dimensions of each layer are shown in Figure~\ref{fig:LANL_I_sys}. 
There is one intentional defect, which is the notch (marked in red in Figure~\ref{fig:LANL_I_sys}). 
The acoustic speed of the materials used in this experiment are 1.5 km/s, 2.82 km/s, and 2.62 km/s for the water, Plexiglas, and concrete layers, respectively. 
Also, the densities of the materials are 997 $\text{kg/m}^3$, 1180 $\text{kg/m}^3$, and 1970 $\text{kg/m}^3$ for the water, Plexiglas, and concrete layers, respectively. 

In this system, there is one well-collimated beam transmitter and 15 receivers mounted vertically. 
The data was collected over a rotational span of $180^\circ$, with a $5^\circ$ step size, for a total of 37 scans.
At the rotational position of $90^\circ$, the sensor assembly is facing the middle of the notch. 
In addition, before each run, the source was tilted upward by a firing angle of $5^\circ$. 
The data was sampled at 2 MHZ. 
Three separate input signals to the system have a central frequency of 29 kHz, 42 kHz, and 58 kHz, respectively. 
Due to space limitations, we will use only the data associated with the 58 kHz signal. 
More detail about the experiment and the designed transducer can be found in~\cite{alanazi2022model,alanazi2022modelX,pantea2019collimated}.

Before testing our method on real data, we simulated two data sets (with and without the notch) using 2D K-Wave simulations  \cite{treeby2010k}. 
The materials used along with their properties are set to the values used in the real experiment. 
The pixel pitch in both the horizontal and vertical axes is set to 1 mm. 
Since the computational domain in K-Wave must be finite, we use a perfectly matched layer (PML) \cite{treeby2010k}. 
In our simulations, we assume that the PML starts from the outer boundary of the computational domain and extends for 20 grid points in all directions. 
The source used in the simulations is a linear-array transmitter, tilted upward by $5^\circ$, and backed with an isolator to prevent the wave from propagating backward, and focused on the forward direction. 
The input signal used is the 58 kHz signal from the experimental data. 

The UMBIR forward and prior models parameters are set to the values reported in \cite{alanazi2022modelX}. 
The parameters of the forward and QGGMRF agents in RARE-MACE are the same values used in UMBIR models. 
The RARE-MACE algorithm is initialized with UMBIR reconstructions to speed up convergence.
The BM3D agent uses the default values in \cite{makinen2020collaborative} and the noise kernel in Figure~\ref{fig:kernel}(c).
The value of the parameter $\rho$ is set to 0.9, $\mu$ is set to 1 to give equal weights for the forward agent and the priors, and the total number of iterations is 100.

\begin{figure}[htb]
\centering
\includegraphics[width=0.4\textwidth]{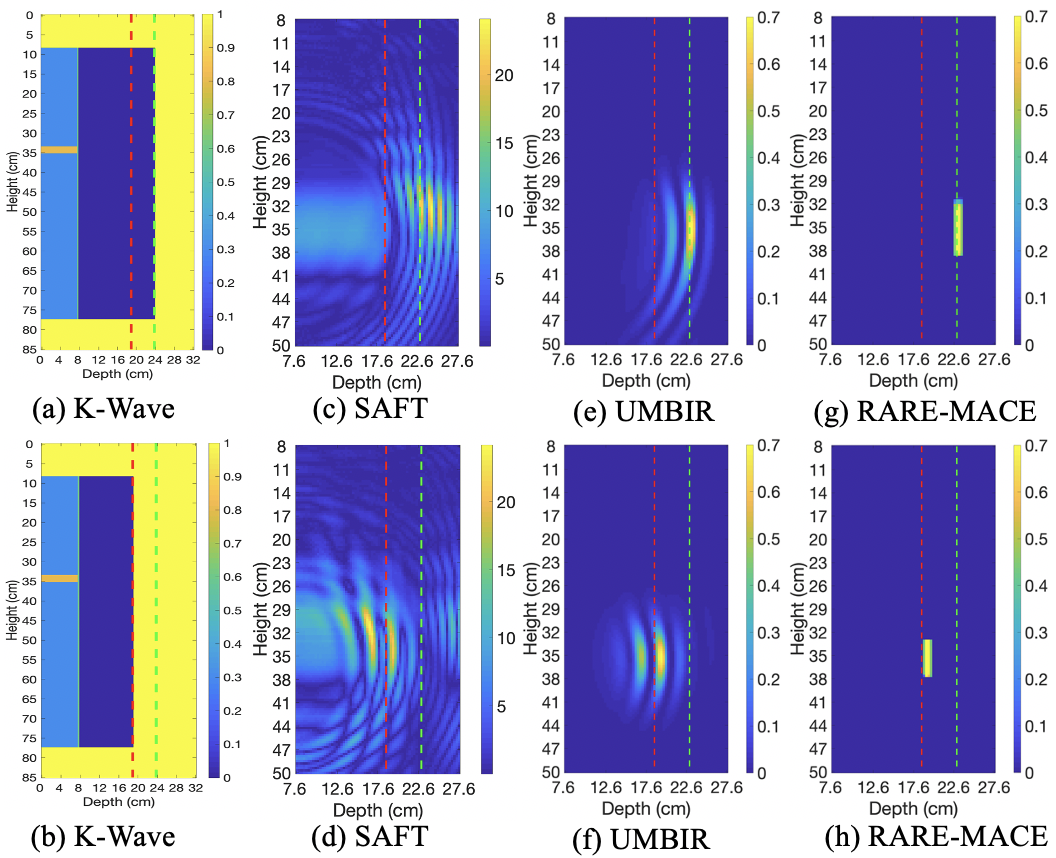}
\caption{(a) The ground truth used in K-Wave for the specimen without any defects and (b) with the notch. (c) and (d) SAFT reconstructions. (e) and (f) UMBIR reconstructions. (g) and (h) RARE-MACE reconstructions. The red and green dashed lines demonstrate the notch and back wall locations, respectively. RARE-MACE substantially suppressed ringing artifacts.}
\label{fig:kwave_results}
\end{figure}
\begin{figure*}[htb]
\centering
\includegraphics[width=0.8\textwidth]{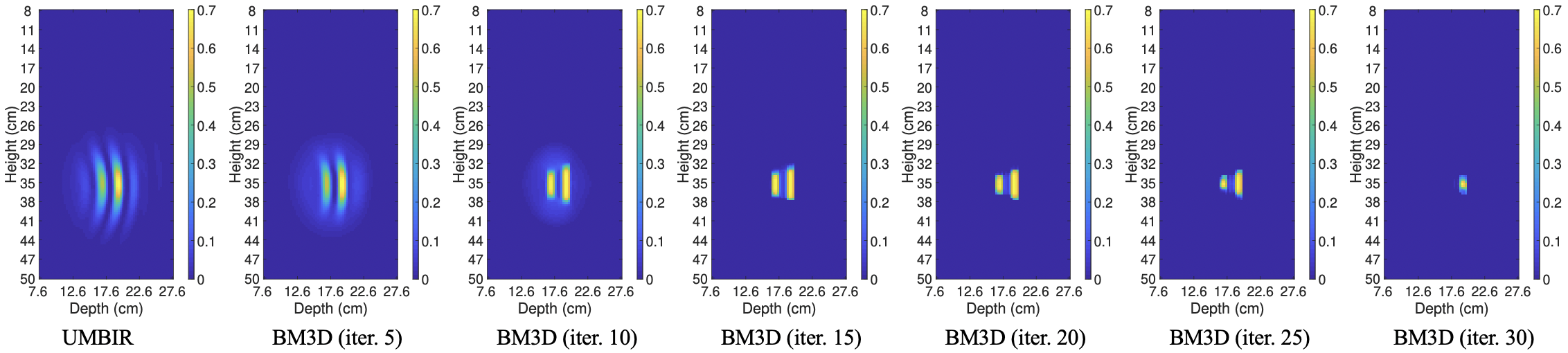}
\caption{
An UMBIR reconstruction using K-Wave data along with repeated applications of BM3D using the correlated noise kernel as a post-processing step.  
The output of every $5^{\text{th}}$ iteration is shown from left to right. 
Iteratively applying BM3D with the circular correlated noise kernel reduces ringing artifacts but also excessively attenuates the true reflection, in contrast to the RARE-MACE results in Figure~\ref{fig:kwave_results}.  
}
\label{fig:impact_bm3d}
\end{figure*}

Figure~\ref{fig:kwave_results} shows a comparison of SAFT, UMBIR, and RARE-MACE when reconstructing synthetically generated concrete cylinder data. 
Figure~\ref{fig:kwave_results}(a) and (b) show the ground truth used to generate the synthetic data both with and without the notch. 
The light blue region is water, dark blue is concrete,
green is Plexiglas, and orange is a vacuum to block direct arrival signals. 
The yellow regions beyond the boundaries are air to mimic the real data. 
Without the notch, the back wall should be located at a depth of 18.85 cm (shown with a red dotted line), and with the notch it should be located at a depth of 23.85 cm (shown with a green dotted line).

Figure~\ref{fig:kwave_results}(c) and (d) show the corresponding SAFT reconstructions.
Notice that that both SAFT reconstructions have considerable uncertainty in the location of the reconstructed back wall, and that both reconstructions have strong ringing artifacts.
In contrast, the UMBIR reconstructions in Figure~\ref{fig:kwave_results}(e) and (f) produce significantly more accurately localized estimations of the notch and back wall locations.
However, UMBIR still has significant ringing artifacts.

The RARE-MACE reconstructions in Figure~\ref{fig:kwave_results}(g) and (h) are significantly better than the UMBIR reconstruction, with substantially suppressed ringing artifacts and a much more accurate reconstruction of back wall and notch, respectively. 
These results demonstrate the significance of our proposed RARE-MACE framework, which can exploit advanced priors such as BM3D with a noise kernel desiged to suppress ringing artifacts induced by fractional bandwidth effects in ultrasound images. 

To provide insight into the role of the BM3D agent, Figure~\ref{fig:impact_bm3d} illustrates the influence of BM3D using the circular noise kernel as a post-processing step on UMBIR reconstructions. Figure~\ref{fig:impact_bm3d} displays the UMBIR reconstruction on the left together with the output of repeated applications of BM3D with the circular noise kernel on this reconstruction, with the output of every fifth iteration shown from left to right. It is evident that iteratively applying BM3D with the specified noise kernel decreases ringing artifacts, but simultaneously greatly attenuates the actual reflections. Therefore, incorporating BM3D alongside a data-fidelity agent and an edge-preserving agent, such as QGGMRF, as in RARE-MACE, leads to better reconstructions.

\begin{figure}[htb]
\centering
\includegraphics[width=0.4\textwidth]{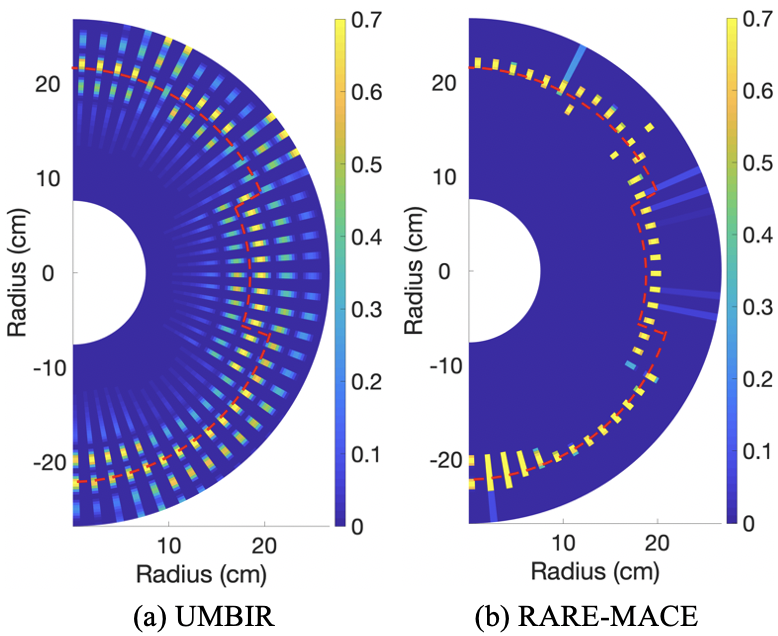}
\caption{(Left) UMBIR and (right) RARE-MACE panoramic stitched reconstructions of the concrete cylinder at a fixed height of 27 cm obtained at each of 37 views in the range from $0^\circ$ to $180^\circ$ and
displayed in polar coordinates. The red dashed line shows the location of the wall and notch, which is closely followed by RARE-MACE reconstruction.}
\label{fig:real_data_results}
\end{figure}

Figure~\ref{fig:real_data_results} depicts a panoramic reconstruction view of the concrete cylinder with measured experimental data using UMBIR and RARE-MACE.
The panoramic reconstruction is formed by combining the views from each measured angle (37 equi-spaced angles from $0^\circ$ to $180^\circ$) to form a 2-dimensional horizontal cross-section at a fixed height of 27 cm. 
In this case, both the back wall with and without the notch is shown as a dotted red line. 
Notice that the UMBIR reconstruction localizes the back wall and notch but with significant ringing artifacts. 
However, the RARE-MACE reconstruction localization closely follows its true location with significantly reduced ringing artifacts. 

\vspace*{-\baselineskip}
\section{Conclusion}
\label{sec:conc}

In this paper, we proposed the RARE-MACE framework, which is designed to suppress ringing artifacts in ultrasound reconstructed images due to the high fractional-bandwidth excitation signal. 
Our method utilizes multiple advanced priors such as BM3D that promote their preferences into the consensus solution of MACE to eliminate various defects such as ringing artifacts in reconstructions. 
Our results using simulated and experimental measurements demonstrated that our RARE-MACE framework shows clear improvements over SAFT and UMBIR and is effective for real data applications.

\vspace*{-\baselineskip}
\section{Acknowledgment}
A. M. Alanazi was supported by King Saud University. 
C. A. Bouman was partially supported by the Showalter Trust and by the U.S. Department of Energy. 
C. A. Bouman and G.T. Buzzard were partially supported by NSF CCF-1763896. 
S. Venkatakrishnan was supported by the U.S. Department of Energy staff office of the Under Secretary for Science and Energy under the Subsurface Technology and Engineering Research, Development, and Demonstration (SubTER) Crosscut program, and the office of Nuclear Energy under the Light Water Reactor Sustainability (LWRS) program.

\bibliographystyle{IEEEbib}
\bibliography{icip2023}

\end{document}